\documentclass[12pt,preprint]{aastex}
\usepackage[]{natbib}
\usepackage{graphics}
\usepackage{amssymb}
\usepackage{amsmath}
\newcommand{\obja}{SDSS J153703+533219}
\newcommand{\objb}{SDSS J145926+493136}
\newcommand{\objc}{SDSS J081102+500724}
\newcommand{\objd}{SDSS J210757-062010}

\begin{document}

\title{XMM observations of BAL Quasars with polar outflows}
\author{Junxian Wang, Peng Jiang, Hongyan Zhou, Tinggui Wang, Xiaobo Dong, and Huiyuan Wang}

\affil{Center for Astrophysics, University of Science and
Technology of China, Hefei, Anhui, P.R.China}
\email{jxw@ustc.edu.cn}

\begin{abstract}

We have selected a sample of broad absorption line (BAL) quasars which show
significant radio variations, indicating the presence of polar BAL outflows.
We obtained snapshot XMM observations of four polar BAL QSOs, to check whether 
strong X-ray absorption, one of the most prominent characteristics of most 
BAL QSOs, also exist in polar outflows.
Two of the sources are detected in X-ray. Spectral fittings show that they
are X-ray normal with no intrinsic X-ray absorption,
suggesting the X-ray shielding gas might be absent in polar BAL 
outflows. Comparing to non-BAL QSOs, one of two X-ray nondetected sources 
remains consistent with X-ray normal, while the other one, which is an
iron low-ionization BAL (FeLoBAL), shows an X-ray 
weakness factor of $>$ 19, suggesting strong intrinsic X-ray absorption. 
Alternative explanations to the nondetection of 
strong X-ray absorption in the two X-ray detected sources are 1) the absorption is more complex than a simple
neutral absorber, such as partial covering absorption or ionized absorption;
2) there might be
significant jet contribution to the detected X-ray emission.
Current data is insufficient to test these possibilities, and
further observations are required to understand the X-ray nature of polar BAL
outflows.
\end{abstract}

\keywords{quasars: general --- quasars: absorption lines
--- radiation: radio continuum}

\section{Introduction}
Broad UV absorption troughs in resonant lines are seen in 
about 10-20 \% optically selected QSOs (Hewett \& Foltz 2003; Reichard et
al. 2003a).
The broad absorption lines (BALs) are found blueshifted up to 0.1c to the
corresponding emission lines, suggesting that they are formed
in a partially ionized wind, outflowing from the quasar. 
The covering factor of the BAL Region (BALR) was found to be $<$ 20\%
based on the 
observed flux ratios of the emission to absorption line
(Hamann, Korista \& Morris 1993). It's found that
the properties of UV
continuum and emission lines of BAL QSOs and non-BAL QSOs are
similar (Weymann et al. 1991; Reichard et al. 2003b).
Observations also show that the de-absorbed X-ray emission of BAL QSOs is 
consistent with non-BAL QSOs (e.g. Green et al.  2001).
Recently, Gallagher et al. (2007) found mid-infrared properties of BAL QSOs
agree with those of non-BAL quasars.
It is now widely accepted that BALR covers only a small fraction of sky, and
most (if not all) BAL QSOs represent a special line of sight toward a QSO 
nucleus (Weymann et al. 1991).

The popular models suggest that BAL quasars are normal quasars viewed
edge-on, skimming the torus and through a wind. For instance, Murray 
et al. (1995) suggested that BALs present
themselves when the line of
sight passes through the accretion disk wind nearly at the
equatorial plane. Elvis (2000) appealed to a funnel-shaped thin
shell outflow that arises from the accretion disk to explain
various observational properties of quasars. 
Both models require a rather large incline
angle of $i\sim 60^{\circ}$ for BAL quasars.
However, recent numerical work indicates that it is also plausible to launch
bipolar outflows from the inner regions of a thin disk (e.g. Punsly 1999;
Proga \& Kallman 2004). There are growing observational evidence indicating
the existence of polar BAL outflows (e.g. Brotherton, De Breuck \& Schaefer 2006; 
Punsly \& Lipari 2005; Ghosh \& Punsly 2007; Zhou et al. 2006; Becker et al. 2000;
Barvainis \& Lonsdale 1997; Jiang \& Wang 2003).

BAL QSOs are found to be X-ray weak following the work of Green \&
Mathur (1996, but see Ghosh \& Punsly 2008 for the discovery of three soft 
X-ray loud BAL QSOs without intrinsic X-ray absorption). For radio quiet BALs, there are strong
evidence showing that the weakness in X-rays is not intrinsic but due to
strong X-ray absorption, with the hydrogen column density of a few
10$^{22}$ to $>$ 10$^{24}$~cm$^{-2}$ (Wang et al. 1999;
Mathur et al. 2000; Gallagher et al. 2002, 2006;
Grupe, Mathur \& Elvis 2003).  
Although UV and X-ray absorption are clearly linked, the X-ray column density
detected in the X-ray band is much higher. A possible model for the 
X-ray absorber is the "shielding gas" postulated by Murray et al. (1995)
and it could be generated naturally in the simulations of Proga, Stone \& Kallman (2000).
Radio loud BALs appear similar in X-ray. Brotherton et al. (2005)
found that similar to radio-quiet BALs, 5 radio loud BALs are also weaker 
in X-ray by factors of 40 or more compared to normal unabsorbed radio loud 
quasars. However, the hardness ratios of these radio-loud BALs are rather
soft, inconsistent with simple neutral absorption. The unabsorbed X-ray
emission might be associated with a radio jet, a partial-covering
absorber, or scattered emission which is not absorbed.
Similarly, Miller et al. (2006) found variable and complex soft X-ray absorption
(probably a partial-covering absorber) in radio-loud BAL Quasar PG 1004+130
(also see Schaefer et al. 2006 for X-ray data of radio-loud BAL Quasar FIRST J101614.3+520916).

In a previous paper, we have selected a sample of BAL QSOs from 
the SDSS (the Sloan Digital Sky Survey, York et al. 2000) Quasar 
Catalog with significant
variations in radio band (Zhou et al. 2006). The large amplitudes
of the variations imply brightness temperatures much higher than
the inverse Compton limits (10$^{12}$ K), suggesting the presence of 
relativistic jets beaming toward the observer.
As revealed by both theoretical models and numerical simulations 
(e.g. Blandford \& Znajek 1977; McKinney \& Gammie 2004), the relativistic 
jet is believed to be perpendicular to the inner accretion disk.
Also note that VLBI observations have shown the inner accretion disk 
(within 1 pc scale) in NGC 1068 is aligned perpendicular to the radio jet 
(Gallimore, Baum \& O'Dea 2004). 
This implies that the BAL outflows in these quasars are almost orthogonal to
the accretion disc. 
We propose snapshot XMM observations to check whether strong X-ray
absorption, one of the most prominent characteristics of most BAL QSOs, also
exist in the polar outflows, and to check whether face-on BAL QSOs are
otherwise X-ray normal.
In this letter we present XMM observations on four of the BAL QSOs with polar
outflows.
Throughout the paper, we assume a cosmology with $H_{0}$= 70 km\,
s$^{-1}$\,Mpc$^{-1}$, $\Omega_{M}=0.3$, and
$\Omega_{\Lambda}=0.7$.

\section{Sample selection, Observations and Data analysis}
In the previous paper (Zhou et al. 2006), we have selected a sample of 8 BAL
QSOs with polar outflows from SDSS DR3. XMM exposures on randomly selected 
four sources were obtained to study their X-ray properties.
We note that Ghosh \& Punsly (2007) applied similar selection to SDSS DR5, and
corrected one error in our calculation of the low limit of the observed radio 
brightness temperature (Zhou et al. 2006). We revised our calculations and 
found three BAL QSOs with XMM exposures have observed brightness temperature ($T_b$) $\gtrsim$
10$^{12.7}$ K, well in excess of the inverse Compton limit, while the rest one,
\objd\, has $T_b$ $\gtrsim$ 10$^{12}$ K.
We note that however, the inverse Compton limit is a very conservative upper limit,
which requires the radiating plasma to be very far out of equipartition
(see Kellermann \& Pauliny-Toth 1969), while the equipartition brightness 
temperature is $\sim$ 5 $\times$ 10$^{10}$ K (Readhead 1994).
Homan et al. (2006) have measured the intrinsic $T_b$ of the parsec-scale jet
cores of AGNs and found intrinsic $T_b$ of $\sim$ 2 $\times$ 10$^{11}$ K in
their maximum brightness states.
Based on these facts, we conclude that it's still safe to consider all the
four sources as polar BALs.

In table 1 we present the observational log for the {\it XMM-Newton} exposures
on four BAL QSOs. During the exposures 
the EPIC-PN, MOS1 and  MOS2 cameras were operated in Full Frame mode and 
the thin filter was utilized.
The Observation Data Files were processed to produce calibrated lists using the
XMM-Newton Science Analysis System (SAS 7.0.0).
Bad pixels were removed and EPIC response matrices were generated using the SAS tasks ARFGEN and RMFGEN. 
Time intervals of high background flaring were eliminated by examining the
light curves of photons above 10 keV.
The total good exposure times for all observations were listed in Table 1.

Two of the sources (\objb\ and \objd) were not detected by any of the
cameras (see Figure 1). For each source we extracted the X-ray photons
from a circular region with a radius responding to the half energy width
(HEW) of the PSFs\footnote{15.2\arcsec, 13.8\arcsec, 13.0\arcsec
respectively for PN, MOS1, MOS2}.
After subtracting the local background and applying aperture corrections,
we calculate 1$\sigma$ upper limit of 2.0 -- 10.0 keV observed fluxes by
assuming a power-law with photon index of 1.9 (see Table 1). 

The rest two sources (\obja\ and \objc) were detected by PN, MOS1 and MOS2
(see Fig. 2). 
We extract the source spectrum for each source from a 24\arcsec
circle, centered at the nominal source position. The bland field background
spectra were extracted from nearby circular region with radius of 40{\arcsec} free
of other sources in the same CCD chip. Single and double events were selected
for the PN detector, and single-quadruple events were selected for the MOS.
We obtained 201.6$\pm$20.8 net counts (PN) in 0.5 -- 10 keV band for \obja\ and
174$\pm$17.8 for \objc, while the expected background level are 98.4$\pm$5.9
and 63$\pm$3.9 respectively.
The resulting spectral files were grouped with at least
one count per bin and C statistic (Cash 1979) was adopted for 
minimization. The
responses of MOS1 and MOS2 cameras are nearly the same, thus we combined their
spectra to increase signal-to-noise ratio (S/N). We fitted the 0.3 -- 10.0 keV 
band spectra of PN and MOS simultaneously (Figure 2).

Our primary model is a power-law with intrinsic neutral absorption. The 
Galactic absorption was also accounted. We find that intrinsic absorption
is statistically not required in either \obja\ or \objc. The fitting 
results are listed in Table 1.
Note all the statistical errors quoted in this paper are for 90\% confidence,
one interesting parameter.
The spectra and the data to model ratios are presented in Fig. 2.

We noticed the very flat X-ray spectrum in \obja\ ($\Gamma$ = 1.31$^{+0.20}_{-0.09}$), which 
could probably be due to more complex absorption.
We thus applied a partial covering fraction absorption model 
instead, and obtained $\Gamma$ = 1.77$^{+0.30}_{-0.36}$, N$_H$ = 
12.9$^{+5.8}_{-8.8}$ $\times$ 10$^{22}$  cm$^{-2}$ and covering factor = 0.64$^{+0.11}_{-0.47}$. 
We also tried an ionized absorption model ($absori$ in $xspec$) to fit the 
spectrum, and obtained $\Gamma$ = 1.47$^{+0.38}_{-0.10}$, N$_H$ =
11.0$^{+27.7}_{-4.2}$ $\times$ 10$^{22}$  cm$^{-2}$, 
the ionization parameter\footnote{$\xi$\ = $L/nR^2$, where $L$ is the 
integrated  incident luminosity between 5 eV and 300 keV, $n$ is the 
density of the material
and $R$ is the distance of the material from the illuminating source
(Done et al. 1992). During the fit, we assumed
a temperature of 3$\times$ 10$^4$ K and solar abundance for the absorber. } 
$\xi$ $\sim$ 2200.
However, extra absorption (either partial covering absorption or warm absorber)
is only statistically required at $\sim$ 97\% confidence level.
For \objc, more complex absorption does not give better fit to the spectrum.

\section{Discussion}
One of the four sources \obja\ was identified as high-ionization BAL
(HiBAL), however, although Al III BAL trough is absent in this source, we can
not rule out the possibility of low-ionization BAL (LoBAL) since Mg II is redshifted out of the
spectroscopic coverage. The rest 3 sources are all LoBAL,
specifically, \objd\ belongs to the rare class of iron LoBAL
(FeLoBAL). This is consistent with the results from Ghosh \& Punsly (2007)
which found that most of the polar BAL QSOs are LoBAL.

Assuming that our BAL QSOs with polar outflows are intrinsically X-ray normal 
comparing with optically selected AGNs, we estimate their expected optical 
to X-ray index $\alpha_{ox}$ using eq. 6 in Strateva et al. (2005). 
We note that using the relation between radio and X-ray luminosities for radio
loud quasars (Brinkmann et al. 2000) yields similar intrinsic $\alpha_{ox}$ 
for these BAL quasars. We can 
clearly see from Table 1 that the two detected sources have 
observed optical to X-ray index $\alpha_{ox}$ well consistent with expected
ones, suggesting they are X-ray normal.
We note the spectrum of \obja\ 
can be marginally better fitted (with a confidence level $\sim$ 
97\%) by a partial covering absorber or a warm absorber.
Adopting the partial covering absorption model, we
obtained an $\alpha_{ox}$ of 1.58 after absorption correction, and for the 
warm absorber model we obtained an $\alpha_{ox}$ of 1.67, both are
still consistent with the expected value 1.69.

Both \objb\ and \objd\ are not detected with the XMM data.
With the derived upper limits of the 2 -- 10 keV fluxes, we obtained lower 
limits to their $\alpha_{ox}$. 
We find that \objb\ is $>$ 2 times fainter in 2 -- 10 keV band than
predicted based on $\alpha_{ox}$ estimation, while \objd\ is $>$ 19 times
fainter. Considering the scatter in the intrinsic $\alpha_{ox}$
of normal AGNs ($\sim$ 0.11, Strateva et al. 2005), the X-ray weakness in
\objb\ is statistically not robust (only at $>$ 1$\sigma$ level), and for \objd\, the X-ray weakness
is at $>$ 3.5$\sigma$ level.
Assuming a powerlaw spectrum ($\Gamma$ = 1.9) with neutral intrinsic absorption,
we estimate the lower limit to the neutral intrinsic absorption (see Table 1). 
Again, due to the large scatter in the $\alpha_{ox}$ estimation, the lower
limit to $N_H$ for \objb\ suffers large uncertainty, and is statistically not robust.

In the popular theoretical model, the BAL outflow, initially launched from
the accretion disk, is accelerated through radiation pressure (e.g. Murray 
et al. 1995). In this model strong X-ray absorption is required as a shield
gas within the outflow, to prevent the UV outflow from being fully ionized.
This is supported by the strong X-ray absorption detected in BAL QSOs.
However, we find no evidence of strong X-ray absorption in two X-ray detected
polar BAL QSOs. Considering the nature of polar outflows might be different
from equatorial outflows, i.e., with different launch places, different
acceleration mechanism (jet might play a major role in the acceleration
of polar outflow), this result is not surprising. If this is true, the
polar BAL might occur at larger distances than equatorial BAL, thus
will not be fully ionized even without thick shielding gas.
A consequence of this hypothesis is that the BAL line profiles might be 
different in polar outflows. Based on the small sample presented in Zhou et 
al. (2006), we can not tell the difference of line profile, line width, etc, if
there is any. A much larger sample would be helpful to test this possibility
by comparing with a well selected control sample. Meanwhile, as Ghosh \& 
Punsly (2007) pointed out, an inordinately large fraction of polar BAL quasars 
are LoBAL, suggesting polar outflows are somehow physically different from
equatorial outflows.

Meanwhile, the X-ray weakness in \objd\ which shows an X-ray weakness factor 
of $>$ 19 suggests the existence of strong X-ray absorption. 
However, we note that \objd\ is an FeLoBAL. There are evidence showing 
that FeLoBAL might signify the transition between an ultraluminous infrared 
galaxy and quasar and does not arise due to a particular line of sight (e.g. 
Farrah et al. 2007). In this case, FeLoBAL might behave differently in X-ray, 
thus its nature does not represent that of polar BAL outflows. Note that
except \objd, all the rest three polar BAL quasars appear consistent with
X-ray normal without strong absorption.

Finally, if strong X-ray absorber does exist in polar BAL quasars,
there are several alternative explanations to the nondetection of X-ray 
absorption in two X-ray detected sources. First, it might 
be due to partial covering X-ray absorption, ionized absorber, or scattering 
component in the spectra, which were found normal in BAL QSOs (e.g. Green et 
al. 2001). In the spectrum of \obja\, we have found that a partial covering
X-ray absorption or ionized absorption could marginally improve the spectral
fitting at $\sim$ 97\% confidence level. For \objc, however
we are unable to confirm or rule out the existence of
such complex absorption. Secondly, there might
be significant contribution from the relativistic jets to the X-ray emission, 
which might originate at larger scale than the shielding gas, and thus be 
free from strong X-ray absorption. This is supported by the very
flat X-ray spectrum ($\Gamma$ = 1.31$^{+0.20}_{-0.09}$) in \obja, which
is normal for radio loud AGNs (but not specifically by the spectrum of \objc).
This is further supported, though not conclusively, by the fact that the observed X-ray luminosities of 
the two detected sources are also consistent with the correlation between 
radio and X-ray luminosities for radio loud quasars (Brinkmann et al. 2000). 
Further studies are required to test above possibilities.

\acknowledgments
The work was supported by Chinese NSF through NSFC10773010, NSFC10533050 and the CAS "Bai Ren" project at University of Science and Technology of China.

\clearpage
\begin{deluxetable}{lccccccccccc}
\tabletypesize{\tiny}
\tablecaption{XMM Observations of four polar BAL quasars\label{tbl-1}}
\tablewidth{0pt}
\tablehead{
\colhead{SDSS J} & \colhead{$z$} &\colhead{type$^b$} &\colhead{Radio} &\colhead{Date} & \colhead{good exp.} & N$_H$ &$\Gamma$ & f$_{2-10keV}$ &C-stat/dof & $\alpha_{ox}$ & $\alpha_{ox}$\\
\colhead{} & & \colhead{} & loudness& & \colhead{ks} &10$^{22}$cm$^{-2}$ & & erg/cm$^{2}$/s& & obs. & exp.
}
\startdata
153703+533219 & 2.4035 & HiBAL? & 31 &2006-06-23 & 8.7 &$<$0.44 &1.31$^{+0.20}_{-0.09}$ & 1.0$\times$10$^{-13}$ &387/474 &1.73 &1.69 \\
145926+493136$^a$ & 2.3700&LoBAL & 29 & 2006-06-09 & 8.5 & $>$4.1 & & $<$1.4$\times$10$^{-14}$ & & $>$1.77 &1.66 \\
081102+500724 & 1.8376& LoBAL & 237 &2007-04-18 & 10.0 & $<$0.59 & 1.71$^{+0.30}_{-0.14}$ &5.1$\times$10$^{-14}$ &289/369 & 1.55 & 1.60 \\
210757-062010$^a$ & 0.6456& FeLoBAL & 72 & 2007-04-24 & 6.3 &$>$40 & & $<$1.0$\times$10$^{-14}$ & & $>$1.93 & 1.54\\
\enddata
\tablenotetext{a}{Not detected by the XMM exposures. Upper limits (1$\sigma$) to
the 2 -- 10 keV flux are given.
The lower limits of absorption column density
(N$_H$) are estimated by comparing the upper limits of X-ray count rates with 
expectations (see \S3 for details). 
}
\tablenotetext{b}{Classified based on most updated SDSS spectrum. 
For SDSS J153703+533219, although Al III BAL trough is absent, we can not 
rule out the possibility of LoBAL since Mg II is redshifted out of the 
spectroscopic coverage.}
\end{deluxetable}

\clearpage
\begin{figure}
\epsscale{1.0} 
\includegraphics[angle=90,scale=0.3]{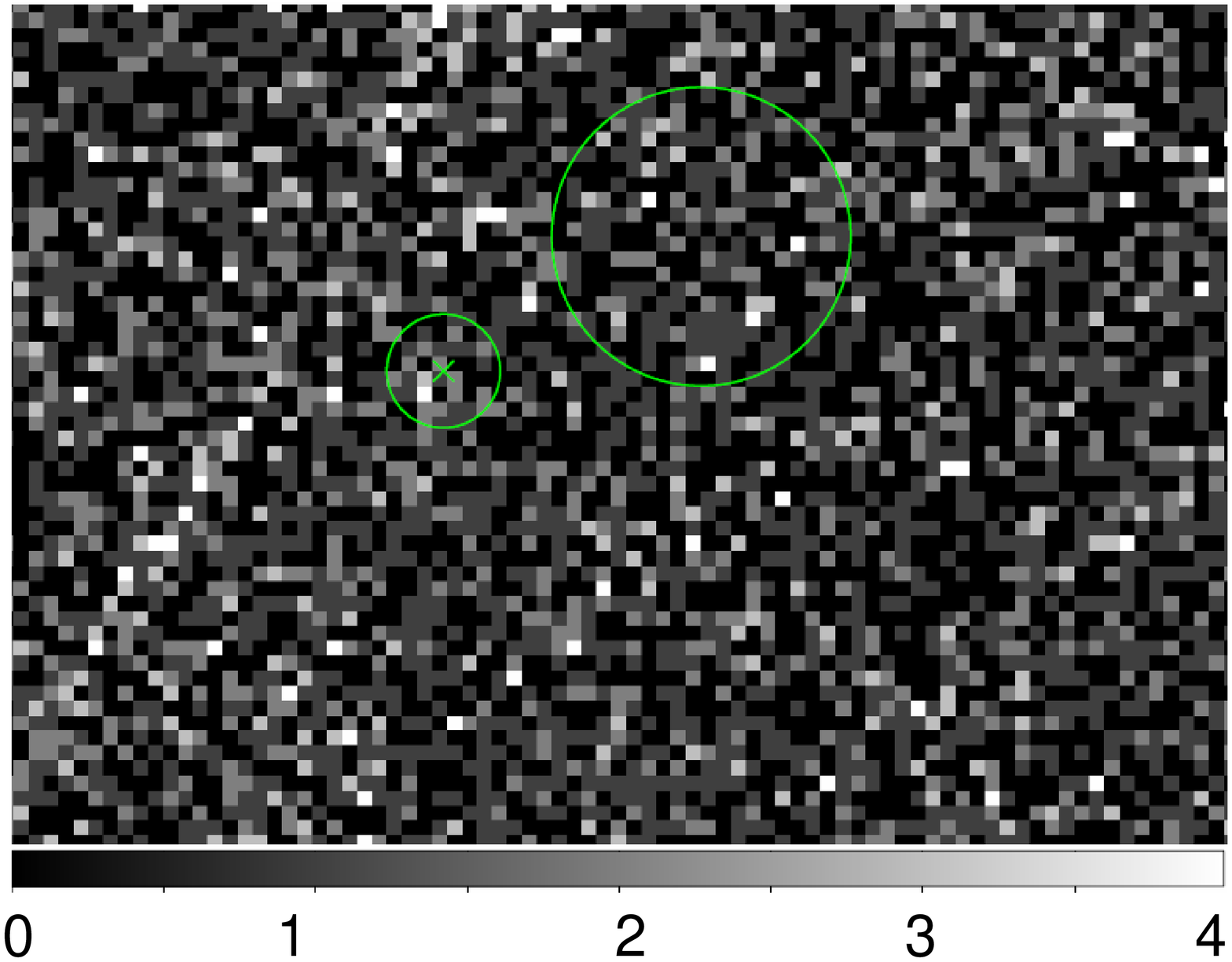}
\includegraphics[angle=90,scale=0.3]{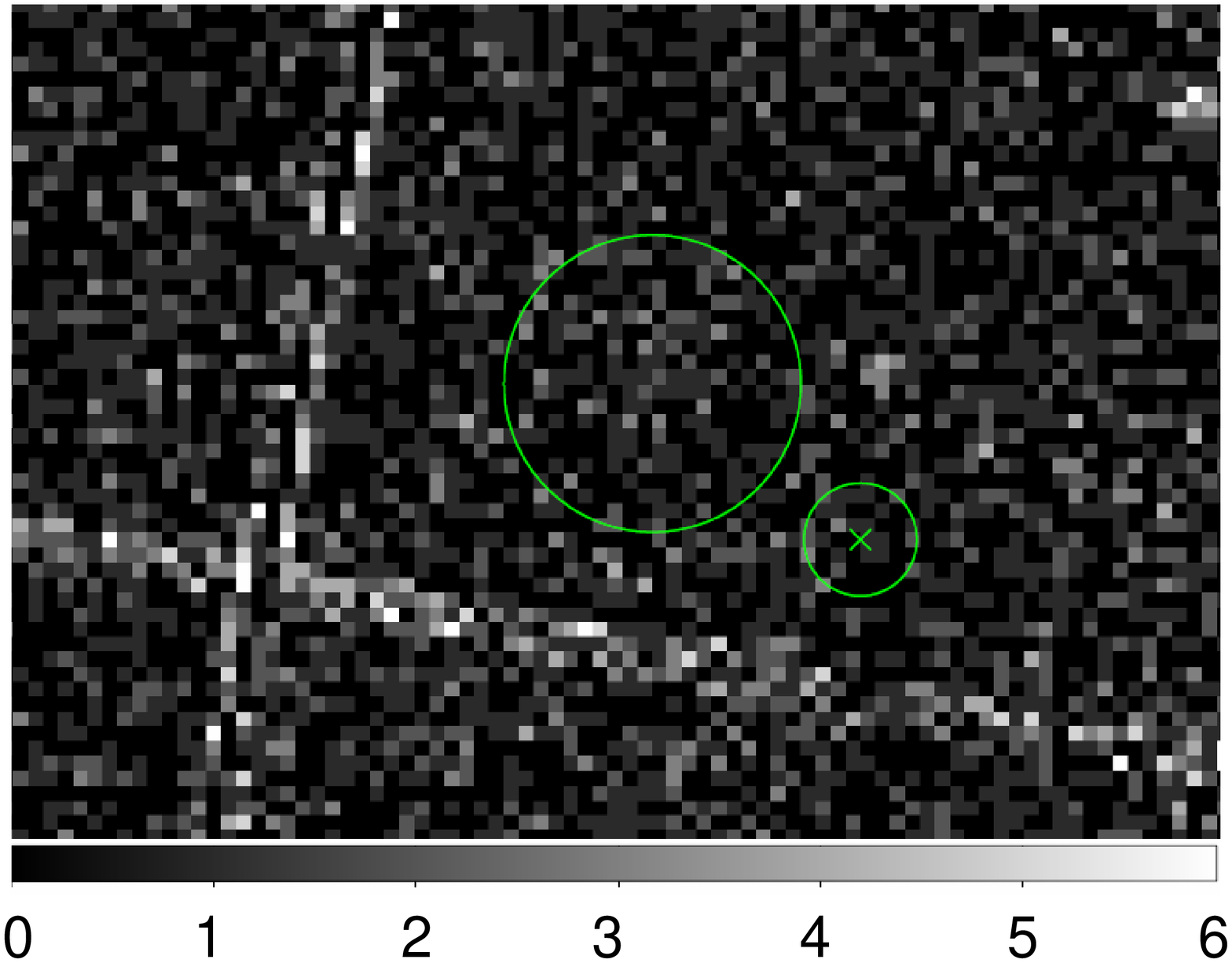}
\caption{XMM PN 0.5 -- 10 keV band images 
(320\arcsec$\times$220\arcsec) of \objb\ (left) and \objd\ (right).
The expected positions of the sources (crosses) on the X-ray images, and 
circles used to extract source (small circles) and background counts (large circles) are marked.
} \label{f1}
\end{figure}

\clearpage
\begin{figure}
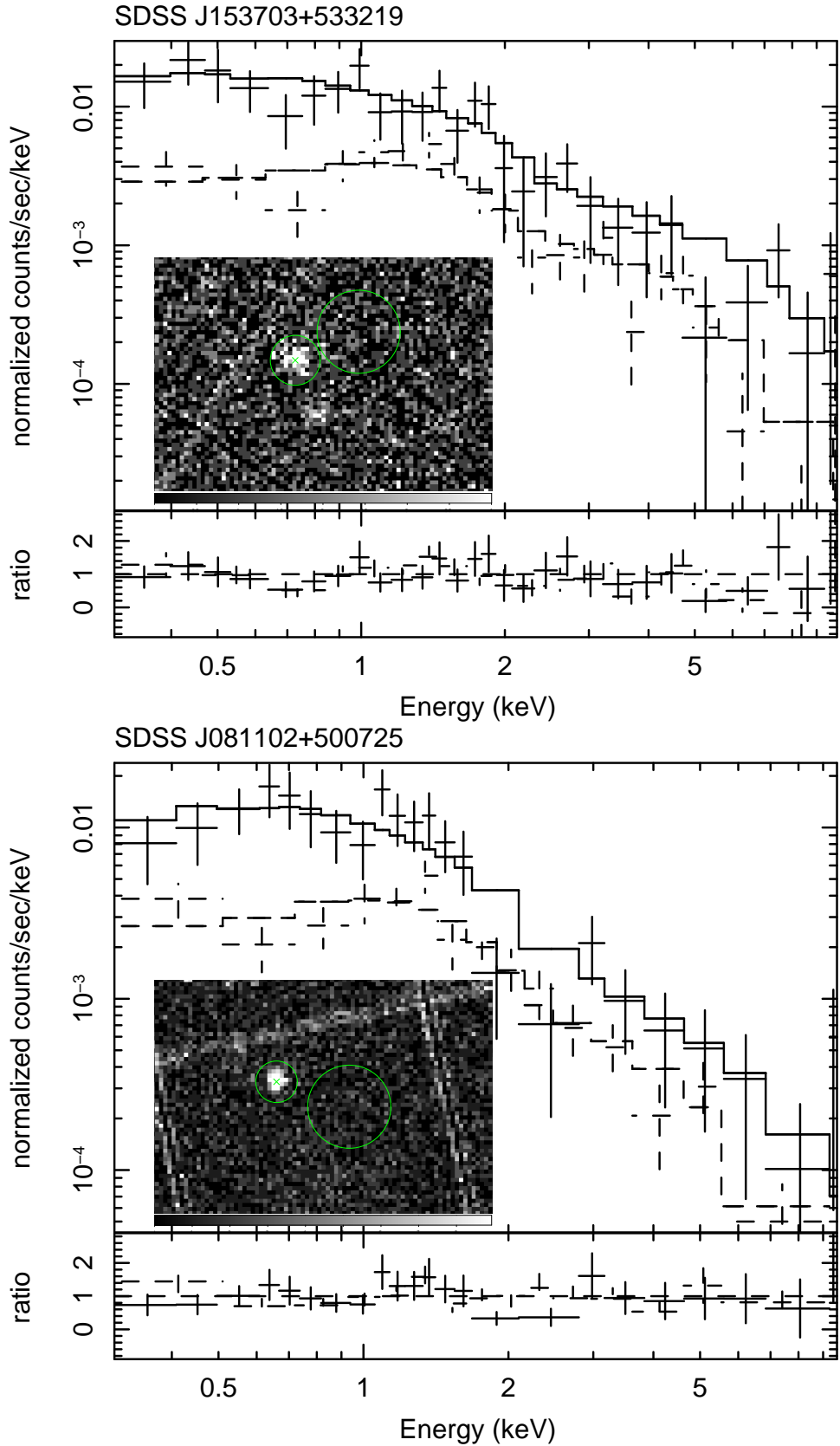

\epsscale{1.0} 
\includegraphics[angle=-90,scale=0.65]{f2a.eps}
\includegraphics[angle=-90,scale=0.65]{f2b.eps}
\caption{Spectral fitting (a powerlaw with neutral absorption) to the XMM PN 
(solid) and MOS (dotted) data of \obja\ and \objb. The cutouts present the
XMM PN 0.5 -- 10 keV band images with source and background regions marked.
} \label{f2}
\end{figure}

\end{document}